# Submillimeter-wave cornea phantom sensing over an extended depth of field with an axicon-generated Bessel beam

Mariangela Baggio, Aleksi Tamminen, Joel Lamberg, Roman Grigorev, Samu-Ville Pälli, Juha Ala-Laurinaho, Irina Nefedova, Jean-Louis Bourges, Sophie X. Deng, Elliott R. Brown, Vincent P. Wallace, *and* Zachary D. Taylor[*]

*Abstract*— The feasibility of a 220 - 330 GHz zero order axicon generated Bessel beam for corneal water content was explored. Simulation and experimental data from the 25-degree cone angle hyperbolic-axicon lens illuminating metallic spherical targets demonstrate a monotonically decreasing, band integrated, backscatter intensity for increasing radius of curvature from 7 – 11 mm, when lens reflector and optical axis are aligned. Further, for radii >= 9.5 mm, maximum signal was obtained with a 1 mm transverse displacement between lens and reflector optical axes arising from spatial correlation between main lobe and out of phase side lobes. Thickness and permittivity parameter estimation experiments were performed on an 8 mm radius of curvature, 1 mm thick fused quartz dome over a 10 mm axial span. Extracted thickness and permittivity varied by less than ~ 25 μm and 0.2 respectively after correction for superluminal velocity. Estimated water permittivity and thickness of water backed gelatin phantoms showed significantly more variation due to a time varying radius of curvature. To the best of our knowledge, this is the first work that describes axicon generated Bessel beam measurements of layered spheres with varying radii of curvature, in the submillimeter range.

## I. Introduction

THz sensing corneal tissue is an expanding research topic that has been investigated by various research groups [1-4]. The key idea is to exploit the high sensitivity of THz waves to water, combined with the well-defined, layered tissue structure, to measure variations of corneal water content. Excessive water content is indicative of edema, which can occur as the result of disease or surgical procedure [5, 6].

Clinical evaluation of THz imaging/sensing of corneal water content assessment has been hampered by numerous practical constraints. One major hurdle is the apparent sensitivity of THz corneal-tissue water content (CTWC) measurements to target location and target motion. Transverse and longitudinal misalignments as small as 0.1 mm and 0.5 mm respectively have been shown to produce extracted water content errors [7]. Data in the literature indicate that patient eyes demonstrate uncontrolled radial displacements as large as 2 degrees during a typical eye exam duration [8] leading to potentially large misalignment error.

Recent work by our group has explored water content assessment via submillimeter-wave reflectometry using a Gaussian-beam telescope and phantoms comprised of gelatin hydrogels on a water backing [7]. Experimental observations confirmed sensitivity to misalignment in both the longitudinal and transverse directions and consequent parameter estimation error. These issues were further explored, theoretically in [7] where the monostatic reception of corneal reflection under Gaussian beam illumination was computed in the 220 – 330 GHz range. Stratified media (SMT) and effective media theory (EMT) were employed with scalar Green's functions to compute the aggregate reflectance spectrum. The received radiation was gauged to a reference reflector and optimization methods were used to extract tissue water content and thickness. This analysis yielded ~120 μm and ~1.2% errors for thickness and water content, respectively, for transverse misalignment as small as 0.5 mm. Similarly, 0.1 mm axial displacement corresponded up to ~92 μm and ~4% for thickness and water content errors, respectively likely due to greater phase error.

For Gaussian beam illumination, the degradation in parameter extraction with misalignment arises from the relatively large beam numerical aperture (NA = $\lambda/\pi w$ = 0.41 @ 275 GHz) and thus a rapidly increasing mismatch between phase front curvature and corneal radius of curvature for non-aligned positions. Axicon-generated Bessel beams (AGBB) [9, 10] are investigated here to counteract misalignment. As indicated by the title, a collimated beam incident on a plano-conical axicon lens creates a good approximation of a zero-order Bessel beam over a long but finite axial region just beyond the tip of the axicon [9, 10]. The wavefront in the Bessel beam region is nearly planar and the transverse extent is constrained due to the approximate non-diffracting nature within the extended focus. Additionally, with larger cone

[*] Mariangela Baggio, Aleksi Tamminen, Joel Lamberg, Roman Grigorev, Samu-Ville Pälli, Juha Ala-Laurinaho, Irina Nefedova, Zachary Taylor are with the Department of Electronics and Nanoengineering, Millilab, Aalto University, Finland (mariangela.baggio@aalto.fi, aleksi.tamminen@aalto.fi, joel.lamberg@aalto.fi, romagrig95@gmail.com, juha.ala-laurinaho@aalto.fi, irina.nefedova@aalto.fi, zachary.taylor@aalto.fi)    .
Jean-Luis Bourges is with the Hôpital Cochin, Université Paris Descartes;

Paris, France (jean-louis.bourges@htd.aphp.fr). Sophie Deng is with University of California, Los Angeles, Department of Ophthalmology, Los Angeles, CA, USA (deng@jsei.ucla.edu). Elliott R. Brown is with Wright State University, Department of Electrical Engineering, Dayton, OH, USA (elliott.brown@wright.edu). Vincent P. Wallace is with University of Western Australia, School of Physics, Maths and Computing, Physics, Perth, Australia (vincent.wallace@uwa.edu.au).

angles, the spot is sufficiently small to approximate the interrogated corneal surface as ~ planar thus enabling application of standard EMT and SMT. These properties make the AGBB a good potential candidate for corneal imaging as axial misalignment should produce limited variations in phase front curvature mismatch (the Bessel beam phase front is planar). Additionally, transverse misalignments should also produce small differences in phase front curvature mismatch since the AGBB central lobe is small (the measured null-to-null widths is between 1.3 -2.4 mm [11]).

Our group has already carried out reflectometry of corneal phantom both with a Gaussian beam telescope (GBT) and partially with the optics presented in this paper. Previously, we extracted the water content of corneal phantoms [12] and quartz permittivity [13], by using the GBT. In [14] quartz permittivity was extracted with an axicon hyperbolic lens, but we did not correct the phase velocity as we do in this paper.

This paper describes a systematic study of using AGBB for standoff permittivity extraction of layered spherical targets in the WR-3.4 (220 – 330 GHz) band. Section II presents a brief review of axicon lenses in the THz regime and details the design of the presented hyperbolic axicon lens. Geometrical optics (GO) and physical optics (PO) simulations of backscatter and coupling efficiency for spherical targets over a range of radii of curvature (RoC) are also presented. An experimental setup and evaluation with fused quartz spherical domes is detailed in section III. Section IV reports results in water content extraction with gelatin hydrogels.

## II. Submillimeter Wave Hyperbolic Axicon Lens Design And Simulation

### A. Background

McLeod proposed the axicon lens in 1954 [15] and it was initially used to generate a focused ring of light. The Bessel beam was first described by Durnin in 1987 [16] and several methods to generate it were subsequently proposed including the axicon. The axicon has already been explored in the THz field in numerous applications including 3d imaging [17] and tomography [18, 19].

Previously our group reported on the design, fabrication and measurement of two hyperbolic-axicon lenses for standoff measurements of spherical targets with 7.5 mm radii of curvature [11]. The two designs incorporated the same hyperbolic first surface for beam collimation but differed by the external cone angle of the axicon; which was either 15° and 25°. The lenses were designed for 220-330 GHz and fabricated of cyclic olefin copolymer (trade name: TOPAS COC). The lens output was evaluated with near-field, mm-wave scanning measurements and confirmed the output field was a close match to a Bessel beam having a radial wave number within 94-96% of the design at frequencies below 280 GHz. Agreement between theory and experiment was further verified from a distance to the tip to 51 and 87 mm. The transverse 1/e beam width was ~ 1 mm and 2 mm for axicon external cone angles of 25 and 15 degrees, respectively. When illuminating a 7.5 mm sphere the 1/e beam edge compared to the beam centroid experiences an additional free space path of 67 μm = 21.6° @ 280 GHz for the 25-degree optic and 271 μm = 87.8° @ 280 GHz for the 15-degree optic. The 25-degree axicon angle was determined more suitable for initial corneal phantom measurement as the reduced beam width is more compatible with standard SMT which assumes planar phase fronts incident on planar targets.

Fig. 1(a) shows a CAD model of the 25-degree hyperbolic axicon lens design and ray tracing indicating the AGBB zone determined by GO. The AGBB axial extent is ~ 58 mm and denoted by $z_{max}$ in Fig. 1(a) and equation (1) where $\omega(f)$ is the frequency dependent beam radius at the axicon surface, $\beta_0$ is the inclination angle of the rays defined by the external cone angle $\alpha$ and lens material permittivity $\varepsilon_r$ as in equations (2) [20]. Equation (3) shows that the axially directed k-vector, $k_z$, is less than the free space k and thus the phase velocity is superluminal [21]; $v_p \sim 1.04c_0$ for $\alpha = 25º$

$$z_{max} = \frac{\omega(f)}{\tan(\beta_0)} \tag{1}$$

$$\beta_0 = \sin^{-1}(\text{Re}\{\sqrt{\varepsilon_r}\}\sin(\alpha)) - \alpha \tag{2}$$

$$k_z = k_0 \cos(\beta_0) \quad \rightarrow \quad v_p(\alpha = 25º) \sim 1.04c_0 \tag{3}$$

The measured beam patterns of the 25-degree axicon at 280 GHz are shown in Fig. 1 (b), (c), and (d) for tip to measurement plane distances of 26, 40, and 50 mm, respectively. The measurements were obtained with a Vector Network Analyzer (VNA), Vector Network Analyzer Extender (VNAX), and near field probe. The location of these measurement planes with respect to the main are indicated by the vertical lines and show that experimental work was performed where the beam was most radially symmetric.

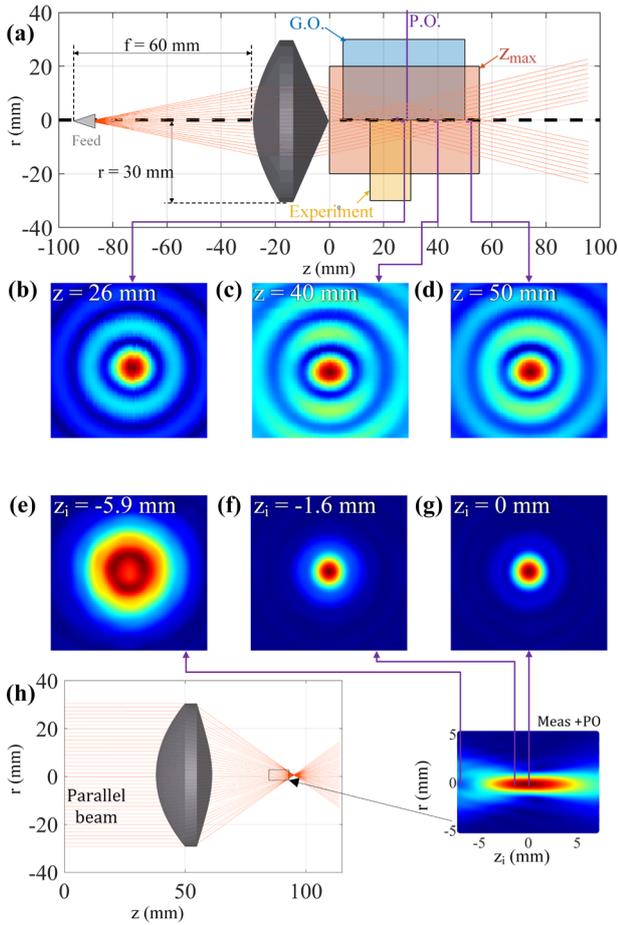

Fig. 1 Map of axial locations with respect to the conical lens tip. (a) The lens creates a Bessel-beam zone over a range [0, $z_{max}$] from the tip, as indicated. The region labeled "GO" shows the range explored via ray tracing. The narrow PO plane at z = 27 mm corresponds to the axial location used in section II.C. The experimental region shown in (a) is the measurement range studied in section III. (b), (c), and (d): beam profiles at 280 GHz obtained at 26, 40, and 50 mm, respectively, from the cone tip in a 10 mm×10 mm area. (h) Schematic of the focusing lens of the Gaussian Telescope from [22]. The zoomed area shows the beam profile near the lens focus. The panels (e), (f) and (g) show the beam amplitude profile in a 10 mm× 10 mm area at the superconfocal, at the subconfocal and the waist location. The beam pattern has been obtained by propagating measured data with PO. In panels (b-g) the amplitude data color scale has been normalized between the local maximum and zero.

The optical parameters of the final lens in our previously reported Gaussian beam telescope are displayed in Fig. 1 (e)-(h). The ray diagram in Fig. 1(h) shows the beam focus ~ 35 mm from the lens second surface apex and the measured beam transverse profile at the beam waist is shown in Fig. 1(g). Physical optics were used to backpropagate the beam to z = -1.6 mm and z = -5.9 mm; corresponding to the subconfocal and superconfocal points respectively [22]. These measurement points occur at approximately the same lens apex standoff as the axicon (~ 25 – 30 mm) and contrast the rapid divergence of the GB over a 5.9 - 1.6 = 4.3 mm span to the relatively small AGBB divergence over a comparatively larger 24 mm span.

B.   *Ray tracing simulation of hyperbolic axicon lens*

A non-sequential ray tracing simulation (OpticStudio/Zemax) was performed to estimate the behavior of the hyperbolic axicon lens, in a monostatic configuration, for varying target RoC and target axial position. The RoC was varied from 7 mm to 11 mm in steps of 0.5 mm for a total of 9 target spheres. The axial position ranged from 5 to 51 mm referenced to the lens conical tip and corresponding to the approximate Bessel beam regime predicted by equation (1).

The simulation distributed 10,000 randomly generated skew rays, subject to a Gaussian pdf, with the source plane located at hyperbolic surface focal point. Input ray positions and departure angles were adjusted to approximate a center frequency of 275 GHz and waist radius of 1.18 mm thus mimicking the corrugated horn antenna beam pattern incorporated into the system described in section III.A.

The dependence of total power collected at the detector to varying walk-off arising from target radius and location required non-sequential ray tracing and ray splitting at each interface to tabulate walk-off losses. However, the low material refractive index (n ~ 1.53) and time-gated post processing used in the experiments of section III.C [12] enabled us to focus on the sequential paths and ignore multipath reflections [8]. The paths were categorized into 4 groups as described in Fig. 2 (a).

Ray tracing simulation results for 4 radii (8, 9, 10, and 11 mm) are shown in Fig. 2 (b). The trends are parameterized by target RoC which is indicated by line style. Path 1 and path 2 are approximately anti-correlated with respect to axial location. The maximum path 1 loss occurs at z = 5 mm and is minimized between 35 – 40 mm for all considered target RoC. Conversely, path 2 loss is minimum at 5 mm and maximum between 45 – 50 mm. In contrast to axial position, path 1 and path 2 are positively correlated in terms of loss as a function of target RoC. More of the remaining energy is lost in path 2 and 3 from the larger RoC spheres than it is from the smaller spheres. In the limit RoC → ∞ (planar target), the loss represented by paths 2 and 3 go to zero and the target reflects all the energy back to the source plane.

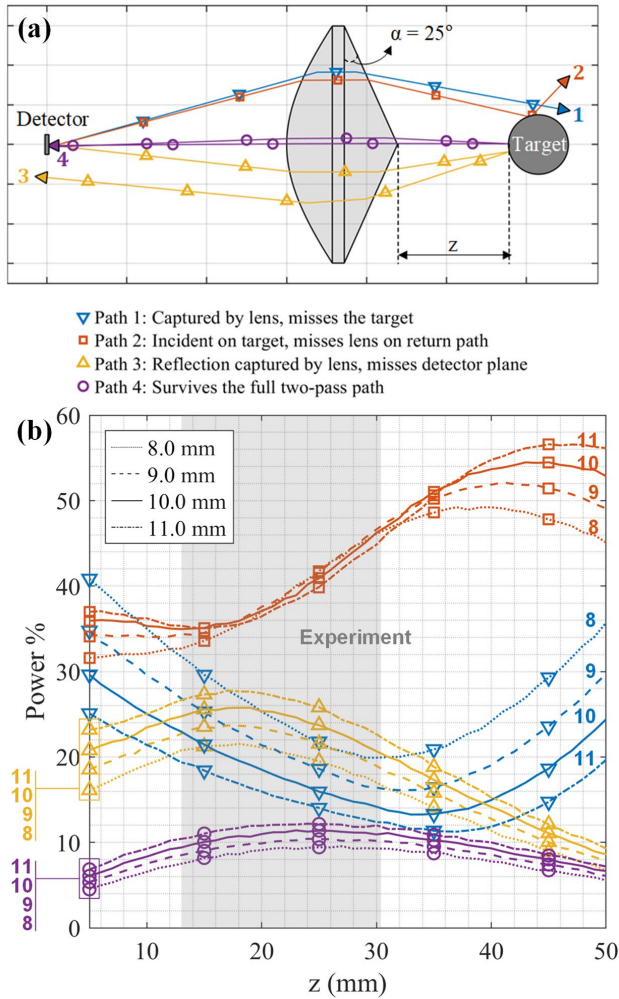

Fig. 2 (a) Non-sequential ray paths in the Zemax simulations. (▽) Captured by the lens but misses the target. (□) Incident on target but misses lens on return path. (△) Back reflection captured by lens but misses detector plane. (○) Survives the full two-pass path. Power-loss percentage of different paths. (b) Non sequential ray tracing power distribution on different path as a function of target distance. The simulation was repeated for sphere of different radii (paths of different colors) and at different positions of its apex. z is the distance between the axicon tip and the sphere apex.

In comparison to the significant variation seen in the paths 1 – 3, the total power collected at the detector plane shows limited dependence on the targets' axial location and RoC. For a given RoC the received power varies by less than 6% over the 45-mm axial range. Further, at a particular axial location, received power varies less than 6% across all tested RoCs. While the ray tracing results suggest sufficient power coupling over a large axial and target curvature range, the analysis shows a non-trivial ray path dependence on RoC; particularly in the walk off losses (paths 2 and 3). The detector plane collected power shows a greater dependence on target curvature than target position and the proportion of power lost to walk-off following scattering from the target is strongly positively correlated with RoC.

### C. Physical optics simulations

Initial analyses backscattered fields and their dependence on target RoC variation transverse misalignment were performed with physical optics with the setups depicted in Fig. 3. PEC spheres with RoC varying from 7 - 11 mm in steps of 0.5 mm were placed individually in front of the axicon with coincident apexes displaced from the cone vertex by 28 mm as displayed in Fig. 3(a). Additionally, the 7 mm and 11 mm PEC spheres were displaced in a direction transverse to the optical axis from 0 to 0.5 mm in steps of 0.1 mm as shown in Fig. 3 (b). For each setup, a scalar Green's function was used to discretize and re-sample the field at the source/detector plane and at each air-dielectric interface to cover the entire sequential path. The source frequency varied in the range of 220 – 330 GHz in steps of 5 GHz and the source field matched corrugated horn [22], modeled as a $TEM_{00}$ Gaussian beam with a spectral dependent waist fit to data extracted from planar near-field measurements. This source was slightly astigmatic, as there is a 0.6-9 % difference in beam waists radii along the two axes [22]. The frequency dependent power coupling between the source distribution and backscattered electric field arriving at the source plane was computed with (4):

$$c_a(f) = \frac{\iint_S E_i(f) \cdot E_r(f) dS}{\iint_S |E_r(f)|^2 \, dS} \qquad (4)$$

where $E_i$ is the co-polar component of the reflected electric field, $E_r$ is the reference (transmitted from the source) electric field and S is the antenna aperture.

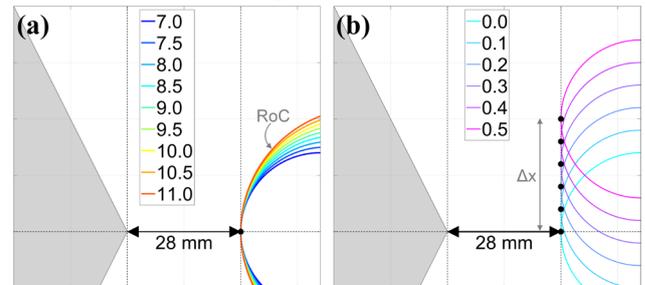

Fig. 3 Geometry for physical optics simulations of target RoC and transverse location. (a) The spherical target apex stays fixed at ~ 28 mm from the conical tip and the RoC is varied from 7.0 mm to 11.0 mm. (b) The target RoC is fixed and the apex is displaced transversely from the beam axis in steps of 0.1 mm to 0.5 mm. Coupling coefficients are computed for each of the perturbations described in (a) and (b).

The coupling efficiency spectrum parameterized by target RoC is shown in Fig. 4(a). The coupling efficiency for target radii from 7 mm – 9 mm feature similar, increasing trends from 220 GHz to a peak at ~ 275 GHz. From 275 GHz to 320 GHz these trends diverge and decrease monotonically with the maximum spread occurring at 330 GHz. The 9.5 mm RoC behaves as a boundary radius for the optical behavior. For radii less than this, the low frequency band displays a shoulder of roughly constant reflectivity from 220 until 275 GHz and then a decrease from 275 to 330 GHz. In contrast, RoCs > 9.5 mm show a monotonic decrease in coupling efficiency across the entire 220 – 330 GHz band. Fig. 4(b) shows the band-integrated reflectivity of each sphere from Fig. 4(a) and demonstrate a clear inverse relation between RoC and coupling efficiency.

Transverse misalignment results are reported in Fig. 4(c). The 7 mm reflectivity spectrum maintains its envelope shape and steadily decreases in total reflected power as displacement from the optical axis is increased. This is further demonstrated by the band integrated signal in Fig. 4(d).

The 11 mm reflectivity spectrum shows a marked difference from the behavior seen with 7 mm. While the behavior vs misalignment at 220 GHz is similar, the response around 275 GHz is approximately invariant to the target position in contrast to the 7 mm sphere which shows substantial spread for the same transverse displacements. Above ~ 275 GHz, the trend is reversed and there is an increase in signal for increasing transverse displacement with the effect enhanced as the frequency is increased.

These results confirm a considerable change in spectral slope for the larger RoC and thus a possible strong confounder to the permittivity extraction as the model-based analysis of aqueous targets depends heavily on the measured slope [12].

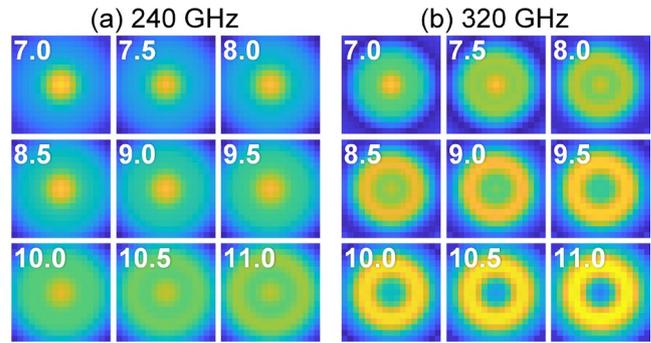

Fig. 5 PO coupling efficiency amplitude at two frequency points (240 GHz and 320 GHz). Each pixel is associated with a different position of the sphere in a 4 mm × 4 mm plane orthogonal to the optical axis. The center position of each panel (0,0) denotes 0 transverse displacement between beam and target optical axis. At 320 GHz, the amplitude of the coupling coefficient is no longer maximized when the target is on-axis for most radii. In each panel, the color scale is between zero and the global maximum.

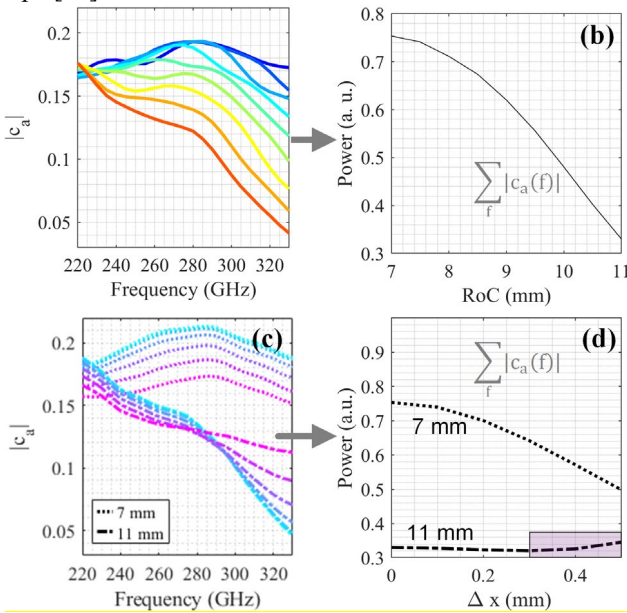

Fig. 4 (a) Coupling coefficient magnitude parametrized by the target radius of curvature of the PEC target. The line color scheme is the same as Fig. 3 (a). (b) Band integrated coupling coefficient as a function of RoC showing a clear monotonic decrease for increasing radius (c) coupling coefficient magnitude of the 7-mm and 11-mm spheres parameterized by transverse displacement. The line color scheme is the same as Fig. 3 (b). (d) band integrated coupling. The 7 mm RoC target coupling shows a clear decrease for increasing misalignment. In contrast, the 11 mm sphere shows a very limited decrease from 0 to 0.3 mm and then a marked increase from 0.3 – 0.5 mm.

A PO simulation of coupling efficiency over a broader 2D space was performed to further explore the strong frequency dependence of the coupling efficiency. The coupling efficiency at 240 GHz and 330 GHz was computed over a 4 mm × 4 mm grid, sampled in 0.25 mm × 0.25 mm steps for a total of 33 x 33 points. The coupling efficiency from all 9 target RoC (from 7 – 11 mm in steps of 0.5 mm) were simulated and the results are presented in Fig. 5 where the simulated RoC is displayed at the top left corner of each panel; the xy span of each panel ±2 mm = 4 mm corresponds to the transverse displacement between beam axis and spherical target axis, and the colormap reports coupling efficiency.

At 240 GHz, maximum coupling efficiency occurs at the center of each panel for all target radii ranging from 7 mm to 11 mm. In other words, maximum coupling at 240 GHz is obtained when the beam and target optical axes are colinear, regardless of target shape. However, a noticeable shoulder starts to develop at 10 mm and transitions to an annular shape with a ~1.2 mm at target RoC 11 mm. Moreover, the annular peak signal is ~ 95% of the center (aligned axes) signal.

For the 320 GHz data, a clear peripheral shoulder coupling coefficient magnitude is already apparent at 7 mm. The shoulder develops into a ~ 1. mm radius ring at 8 mm whose magnitude exceeds the center magnitude at 8.5 mm. Thus, the peak coupling efficiency at 320 GHz is obtained when the beam and target axes are ~ 1 mm displaced and the peak continues to increase, while the center (aligned coupling efficiency) monotonically decreases as the target RoC is increased to 11 mm.

III. MEASUREMENT OF METALLIC SPHERES AND A DIELECTRIC SPHERICAL SHELL

A. Setup

A 50-GHz VNA (Keysight N5225A) with a WR-3.4 (220 – 330 GHz) frequency extender and a corrugated horn antenna [22] was used as the transceiver. The corrugated horn transmit antenna illuminated the 60 mm clear aperture axicon-hyperbolic lens with a Gaussian beam thus generating the quasi-Bessel beam. The backscattered reflection was collected by the lens and coupled back to the horn antenna and VNA extender. Two types of targets were measured: (1) metallic calibration spheres, (2) a fused quartz dome. Each target was translated by high precision linear stages (Physik Instrumente, 1 μm bidirectional repeatability). A swept source - optical coherence tomography (SS-OCT) system was used to collect multiple B-scans of the targets for RoC and thickness verification.

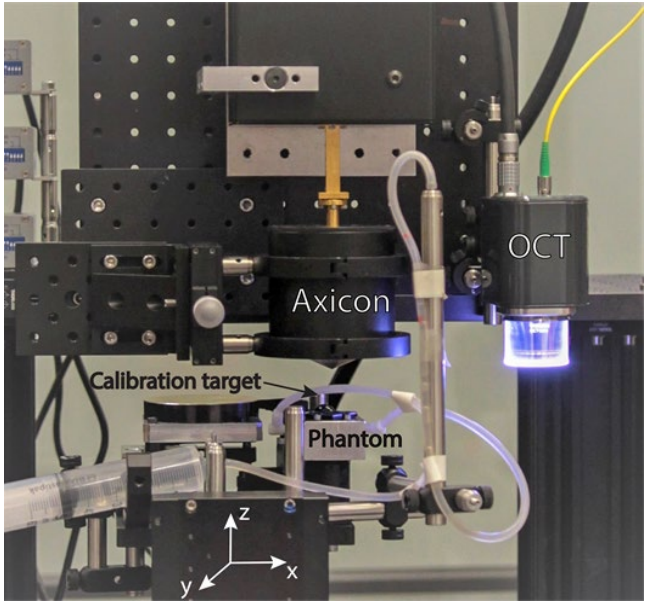

**Fig. 6 Measurement setup:** Each of the three types of targets are placed on precision stages that allow movement in three directions. They can be aligned with the axicon objective or the OCT.

A photograph of the measurement platform beneath the axicon and OCT aperture is shown in Fig. 6. The measurement platform provided a mount for the metallic calibration spheres, reference planar metallic reflector, and phantom holder consisting of an artificial anterior chamber and tubing to maintain pressure.

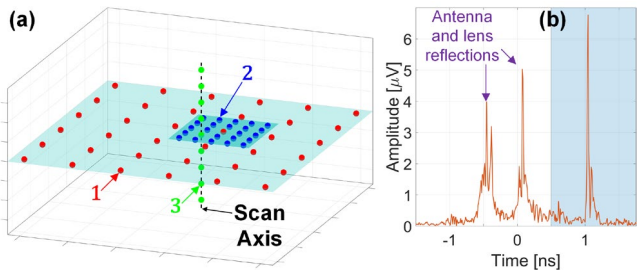

**Fig. 7 Scanning geometry for each target.** First the target is scanned over a coarse grid to locate the approximate transverse location of the maximum reflectivity. A subsequent transverse scan over a finer grid identifies the vertical axis which is scanned at the fine grid step size.

Fig. 7 describes the process to locating a vertical axis along which the targets were scanned. First a coarse grid of points was obtained about an approximate pre-defined location. The grid covered a 4 mm by 4 mm area scanned in steps of 0.5 mm for a total of 89 points. The scattering parameter $S_{11}$ was collected at each location as a function of frequency from 220 to 330 GHz in steps of 13.75 GHz for a total of 8001 points. The spectra were then transformed to the time domain and the maximum (peak) of the time domain representation was located. The peak is proportional to the band integrated $S_{11}$ in the frequency domain and consideration of the peak removes influence of reflections from the horn antenna and backscatter from the lens' hyperbolic and conical surfaces. The spatial location of the maximum reflectivity was identified and set as the center of a fine sampling grid of 1 mm × 1 mm at a step size of 0.1 mm, yielding 121 total points. The time domain peak signal, in the relevant time window, was again computed and the location of the maximum set as the vertical axis transverse coordinates: Targets were scanned along the z-axis for 100 mm-150 mm, depending on the target, with a step of 0.1 mm.

### B.  Metallic reference spheres

Shown in Fig. 8. (a– i) are false color, 3D surface plots of time-domain reflection profiles from metallic sphere (calibration) targets having RoCs between 7.5 and 11 mm. The surface amplitude/color is the time domain pulse peak, which is proportional to the total reflected power within the WR-3.4 band. The course grid values are represented by faded color map and the fine grid values are overlaid on the course grid data.

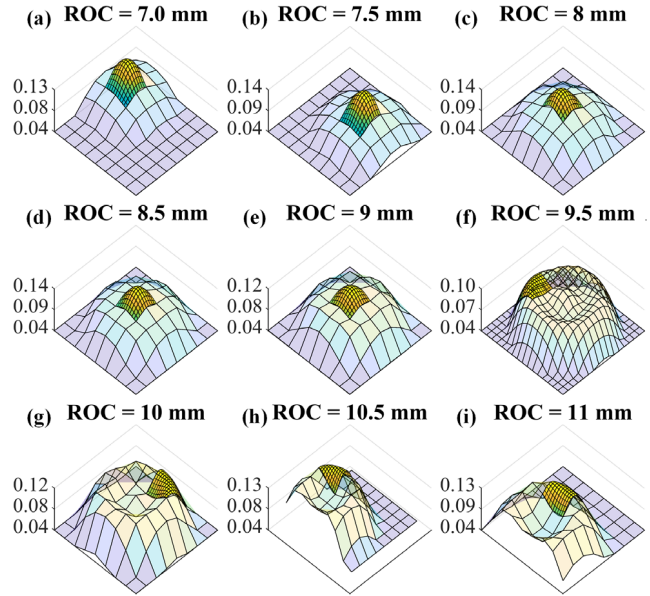

**Fig. 8 Time domain reflection-peak profiles for metal-sphere calibration targets** obtained by raster scans in the x-y plane. The first scan is with a step of 0.5 mm and a range of 4 mm × 4 mm. The second scan, centered on the maximum of the first scan, has a smaller step (0.1 mm) and a smaller range (1 mm × 1 mm). Panel (h) has a scan with step of .25 mm and a range of 4mm × 4mm. The radii of curvature are (a) 7.0 mm, (b) 7.5 mm, (c) 8.0 mm (d) 8.5 mm (e) 9.0 mm, (f) 9.5 mm, (g) 10.0 mm, (h) 10.5 mm, and (i) 11.0 mm. There is a ring of local-maxima points around the axis-of-symmetry when the RoC increases.

The profile for the 7.5 mm RoC sphere shows a clear global maximum surrounded by a low, but discernible reflectivity shoulder. As the target RoC is increased from 8.0 to 8.5 and 9.0 mm, the peak maintains its shape/curvature while the shoulder rapidly rises in amplitude. At 9.5 mm the central peak and surrounding shoulder combine to form a rough flat top. For RoCs of 10.0 mm, 10.5 mm, and 11.0 mm, they form a caldera where the reflection at center is a local minimum surrounded by a ring of peak amplitude. The amplitude peak and ring radius shows limited dependence on RoC for the 10.0, 10.5, and 11.0 mm values.

This behavior is consistent with the PO results in Fig. 4 and Fig. 5 since Fig. 8 reveals a small decrease in 220 GHz reflection concomitant with a larger increase in 330 GHz reflection, as the 11.0 mm sphere optical axis is displaced transversely from the axicon optical axis. Similarly, the transverse misalignment simulations revealed a consistent center amplitude and emerging shoulder at 220 GHz for increasing target RoC, while the 330 GHz data showed the same emerging shoulder combined with a rapidly decreasing center reflectivity. Our results are also consistent with Ref. [23] which studied the vector electric field of an AGBB at 650 GHz. This paper showed that the side lobes have phase opposite to the main lobe. Therefore, presumably they interfere destructively at the receiver when reflected from a sphere. This destructive interference would also explain why the maximum amplitude of the mirror reflection is in a different position from the sphere, as the side lobes tend to increase along the z-axis. This combined with [11]Fig. 1, where one can see that the Bessel beam side lobes at 330 GHz are close to the main lobe, would explain why spherical targets have such a frequency response. It is still not clear whether the caldera is the result of side lobes matching the phase front as the peak tends to occur when the sphere apex is located in-between the main and the side lobes. Either there is no more destructive behavior as the sphere is decentered, or the sphere radius matches the side lobe when the sphere is off axis.

### C. Data processing

Plane wave based analysis in the form of EMT and SMT benefits from calibration of target reflectivity with reflection from reference targets of the same radius and range with respect to the objective optic [12]. The fused quartz domes were scanned over 10 mm and the metallic spheres over 15 mm of axial range with a step of 0.1 mm. Successful calibration of each target reflection coefficient along the z-scan was performed by identifying a calibration target reflectivity from the same axial position with respect to the objective. This was achieved comparing the time domain representations of the calibration and phantom target S-parameters. Each phantom $S_{1,1}$ was compared with the ensemble of S-parameters collected along the sweep axis described in section A for the RoC that most closely matches the target RoC. This process is described by minimizing the sum of square error objective function in equation (5) by finding the optimal z-location of the appropriately sized metallic sphere as in (6):

$$P(z_p, z_c) = \sum_t |\hat{S}_p(t, z_p) - \hat{S}_c(t, z_c, R)|^2 \quad (5)$$

$$z_c^*(z_p) = \underset{z_c}{\mathrm{argmin}} P(z_p, z_c) \quad (6)$$

In equations (5) and (6), P is the objective function, S is the gated $S_{11}$-parameter obtained from the VNA where the time (t) and frequency (f) relations are described by the Fourier transform: $\hat{S}(\cdot) = I^{-1}\{S(\cdot)\}$. The axial locations are denoted by $z_p$ and $z_c$ for the phantom target and calibration reflector respectively. Time domain gating was applied to filter out the clutter, and each $\hat{S}$ was normalized by its peak-to isolate pulse shape and location. The correct calibration position $z_c^*$ for each phantom position $z_p$ was identified by minimizing (5).

### D. Quartz dome analysis

Permittivity and physical thickness extraction were first attempted with a quartz (fused silica) dome (hemispherical shell) procured from VY Optoelectronics Co, Ltd. The dome had an outer radius of curvature of 8 mm and a thickness of 1 mm (Fig. 9(a)). The tolerance for both parameters was 0.1 mm. OCT images of the dome were acquired to verify radius of curvature and central thickness as shown in Fig. 9(b). The optical-to-physical thickness conversion was informed by the refractive index extracted from OCT B-scan measurements of a fused-quartz flat provided by the dome manufacturer. The dome was mounted on the target plate and measured under the axicon objective for a z-scan range of 10 mm.

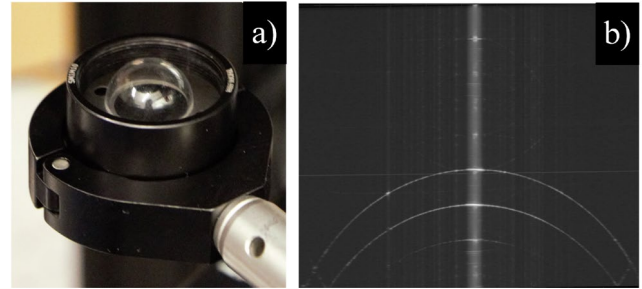

Fig. 9 (a) Fused quartz dome photograph. (b) OCT scan of the quartz dome.

The calibrated reflection coefficient at each point $z_p$ along the axis was computed using the relation in equation (7) where $S_p(f, z_p)$ is the gated, measured signal of the dome and $S_c(f, z_c^*(z_p), R)$ is the correct gated, measured signal of the calibration targets identified via the objective function minimization.

$$\Gamma(f, z_p) = \frac{S_p(f, z_p)}{S_c(f, z_c^*(z_p), R)} \quad (7)$$

$$\Gamma_i = \frac{\rho_i + \Gamma_{i+1} e^{-2jk_i l_i}}{1 + \rho_i \Gamma_{i+1} e^{-jk_i l_i}} \quad \text{for} \quad i = 1,..,M \quad (8)$$

$$\Gamma_{M+1} = \rho_{m+1} \quad (9)$$

$$\rho_i = \frac{n_{i-1} - n_i}{n_{i-1} + n_i} \quad (10)$$

Fitting was performed with particle swarm optimization (PSO) via multilayer calculations [24] described by equations (8–10). In the case of the quartz dome, M = 1 for the quartz layer sandwiched by two air half-spaces. The quartz dome physical thickness $\ell_1$ and refractive index $n_1$ were treated as free parameters in the PSO routines to identify the ($\ell_1$, $n_1$) pair that yields the best fit between $\Gamma_1$ (model) and $\Gamma(f, z_p)$ (measured). Thickness was added as a free parameter to explore if the system SNR was sufficient to constrain optical path length ambiguities [25]. Additionally, to explore the effects of a RoC mismatch between the quartz dome and calibration targets, the PSO fitting was run using data from all the measured metallic spheres with R< 9.5 mm (R = 7.0, 7.5, 8.0, 8.5, 8.7, and 9.0 mm).

### E. Experimental estimates of phase velocity

Calibration measurements with a planar, metallic reflector show evidence of an apparent superluminal beam propagation. Target reflection was acquired at a standoff of ~ 25 mm and then was positioned 15 mm away using high resolution translation stages. The elapsed time between peaks in the Fourier transformed spectra indicate a detected distance of 14.64 mm assuming a phase velocity of $c_0$ suggesting an apparent phase velocity for $(15.0/14.64) \cdot c_0 \sim 1.02 \cdot c_0$. This superluminal phase velocity is consistent with those observed in previous work [21] and, as reported in [21], we assume the velocity is frequency invariant. Therefore, the wavevector in equation (8) becomes $k_i = n_i \omega / 1.02 c_0$

### F. Quartz dome results

Example fits to two different scenarios are shown in Fig. 10. The left column shows result of calibrating the fused-quartz $S_{11}$ with the steel sphere $S_{11}$ of the same RoC (8.0 mm) representing the best-case scenario fits. The right column data analysis was obtained by normalizing the quartz data by the 9.0 mm RoC steel sphere; the largest for which the transverse scan data has a clear central peak (Fig. 7).

Panels (a) and (b) contain the calibrated data in a dotted line style superimposed by the fits described by the extracted thickness d, and refractive index n input into Eqn (8). The curves are parameterized by distance from the ~ start of axicon tip and indicated by the jet color map (blue → 13.1 + 0.0 mm … red → 13.1 + 15.0 mm). Both the 8.0 mm and 9.0 mm steel sphere calibrated data sets show oscillations throughout the band that become more pronounced for frequencies above ~ 275 GHz. These oscillations are due to imperfect time gating. The optical path length between the lens hyperbolic surface and phantom target location is small (compared to [12]) and the bandwidth available at WR-3.4 means the tails of these reflections, in the time domain, overlap. Both results show a sinusoidal envelope characteristic of the reflection from a single layer etalon (air-quartz-air). The 9 mm calibrated results depart from the best fit curves at higher frequencies due to the sensitivity of this sub-band to radius of curvature as discussed in Section II.C and Fig. 3.

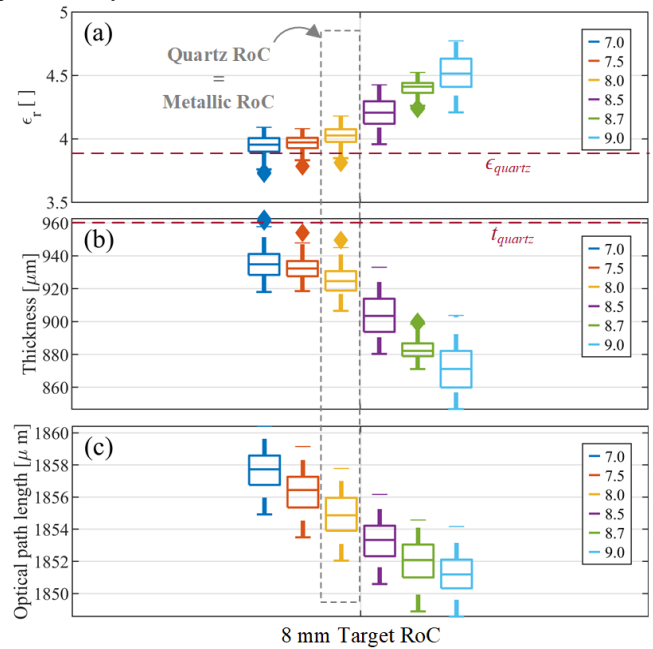

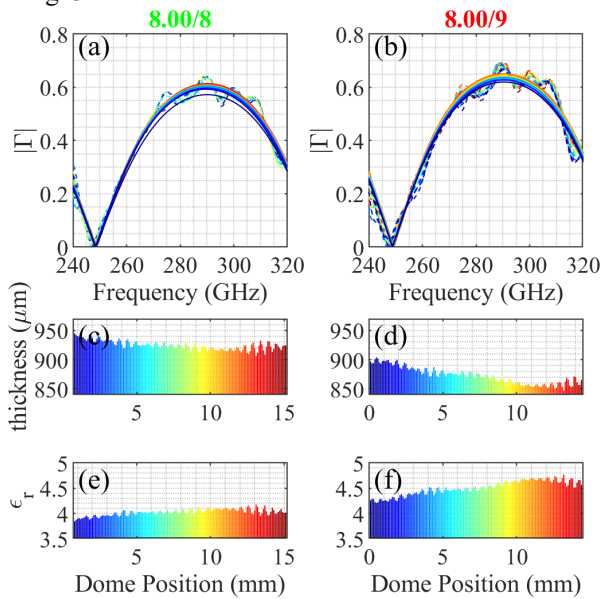

**Fig. 10** Fifteen measurements (dashed lines) and the corresponding estimated reflectivities (solid lines) when the calibration ROC is (a) 8.0 mm (b) 9.0 mm. (c) The estimated thickness and (e) permittivity when the calibration is 8.0 mm. (d) The estimated thickness and (f) permittivity when the calibration ROC is 9.0 mm.

Fig. 10 (c,d) and (e,f) report extracted thickness and permittivity, respectively, as a function of axial location. The extracted data displays a variation vs dome position with an approximate period between ~ 0.25 mm and 0.3 mm. When the extracted permittivity increases, the extracted thickness decreases, thus suggesting a conservation in optical path length. The ensemble permittivity and thickness statistics are $4.02 \pm 0.7$ and $0.925 \pm 0.008$ mm respectively for the 8 mm steel sphere calibration. For the 9 mm steel sphere calibration set, the permittivity increases to $4.50 \pm 0.09$ and the thickness decreases in kind to and $0.872 \pm 0.14$ mm. The thickness extracted with OCT was $0.96 \pm 0.03$ mm and the estimated THz permittivity is ~3.81 [26, 27] with negligible loss. In comparison, the Gaussian beam telescope data resulted in a permittivity of ~3.5 and a thickness of 0.97 mm.

**Fig. 11** Box plots of the quartz dome (a) permittivity, (b) thickness and (c) optical path length for different calibration target RoCs. The box plots show the first, median and third quartile as lines in the boxes. The whiskers are the minima and maxima, and the diamonds are suspected outliers.

Fig. 11 provides a report of the axial location and calibration RoC dependent, extracted quartz dome parameter statistics. The extracted parameters are grouped by the steel sphere RoC (6 groups) and plotted in order of increased RoC. Group statistics (median, interquartile range, etc.) were computed and are reported in the box plots. Data points beyond the whisker bounds are considered outliers. The median of each group is represented by the horizontal line within the boundaries of each box describing the IQR.

The calibration RoC groups are indicated by their color scheme. The extracted permittivity (Fig. 11(a)) shows a clear increase as a function of calibration target RoC and outlier values, above the top of each associated whisker boundary, are observed at 7.0 mm, 7.5 mm, and 8.0 mm. The skewness is most apparent for 7.0 mm and 8.7 mm and the Q1, Q3, and IQR is the smallest for the 8.0-, 8.5-, and 9.0-mm steel spheres. Conversely, the thickness plots (Fig. 11(b)) show a clear decrease for increasing calibration RoC. Outliers are also seen for the 7.5 mm, 8.0 mm, and 9.0 mm groups but they are located below the bottom whiskers.

Fig. 11(c) reports the statistics of the optical path length (OPL), namely the product of extracted thickness and refractive index. The extracted OPL further supports the invariant optical path observation seen in Fig. 10(c-f). The group IQR ranges from 3 μm – 5 μm and the total span, from smallest to largest whisker is ~ 123 μm. There are no outliers in the 7.5 mm, 8.5 mm, and 9.0 mm OPL group, but an outlier has emerged in the 8.0 mm data which represents the best available match between the quartz dome and steel sphere RoC.

IV. PHANTOM TARGETS EXPERIMENTAL RESULTS

Phantoms were fabricated using an injection mold-based process similar to [12]. Improved injection molds were designed and machined. The mold comprised of two halves, one with a 7.8 mm radius of curvature spherical depression and the second half a spherical dome with an inner radius of curvature of either 7.8 – 0.5 = 7.3 mm or 7.8 – 0.6 = 7.2 mm corresponding to spherical concentric shells with 0.5 mm and 0.6 mm thicknesses respectively. The spherical depressions were inset into a circular extruded cut fed by thin channels that connect this compound cavity to the mold's edges as seen in Fig. 12(a). The two halves were mated with steel dowel pins that insure the mold surface concentricity.

thickness were obtained with OCT scans. In Table I the extracted parameters of the selected phantoms are reported. The OCT thickness agrees with the fabrication as the largest deviation from the fabrication value is 25 µm. On the other hand, the RoC varies significantly from the fabrication value, as the radii are increasing during the measurements as already illustrated in Fig. 13. The phantoms frequency sweeps were calibrated with the procedure illustrated in section III.

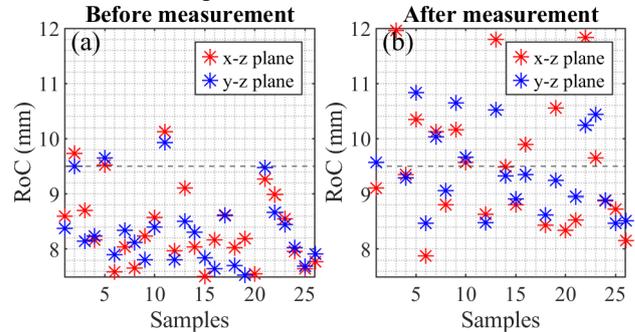

Fig. 13 The RoCs evolution during the measurements. The radii for each sample were measured (a) before and (b) after the MMW measurement.

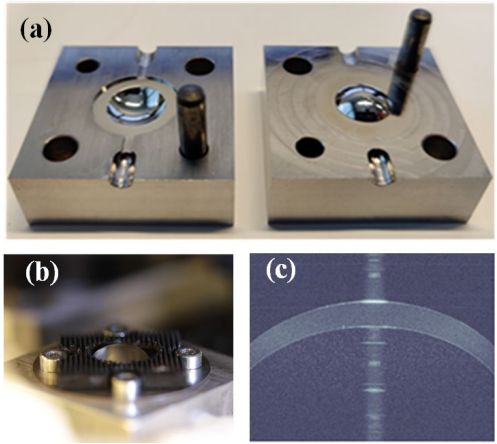

Fig. 12 (a) Open phantom steel mold. The mold is made of two steel blocks. Two dowel pins are used to line up the two blocks together and two screws are used to lock them. (b) Phantom mounted on the AAC (c) OCT image of the phantom. In the image the typical central saturation artifact is present. Underneath the phantom the water is also visible.

Phantoms were fabricated by injecting hot, liquid gelatin into the mold via a syringe. Gelatin is pushed through one channel until it flows out the other channel ensuring a complete filling of the spherical shell cavity. The gelatin phantom was allowed to cool to room temperature and released by separating the halves. The phantoms were mounted to an artificial anterior chamber (AAC) and backed with water at physiologic pressure as described by [12]. Fig. 12 (b) shows the phantom mounted on the ACC, and (c) shows a phantom OCT image.

TABLE. I
PHANTOM THICKNESS AND RADIUS OF CURVATURE

| Phantom number | Mold thickness (µm) | OCT Phantom thickness (µm) | Phantom RoC (mm) |
|---|---|---|---|
| 6 | 600 | 625 | 8.17 |
| 7 | 500 | 517 | 9.14 |
| 8 | 600 | 592 | 8.93 |
| 12 | 500 | 508 | 8.55 |
| 15 | 600 | 611 | 8.86 |
| 18 | 600 | 594 | 8.53 |
| 20 | 500 | 511 | 7.84 |
| 21 | 600 | 591 | 8.75 |
| 24 | 600 | 617 | 8.88 |
| 25 | 500 | 508 | 8.60 |
| 26 | 600 | 600 | 8.33 |

Twenty-six phantoms were measured. Of these, eleven phantoms kept their radii under 9.5 mm during the measurement. We chose to analyze these eleven ones, as the others suffered of the problem illustrated in the previous section. The phantom's RoC, as well as an estimation of the

Two examples of phantoms parameter extraction are shown in Fig. 14. On the left-hand side (a,c,e,g), phantom 25 extraction is reported when the phantom frequency sweeps have been normalized with the frequency sweeps of the sphere of radius 8.5 mm. One can see that the reflection coefficient decreases significantly with increasing frequency. This corroborates the finding in the previous section that suggest that the higher radii have a steeper trend in this frequency band (9.0 vs 7.0 mm). The phantom extracted thickness oscillates between 520 µm and 545 µm. The water content however is about 55 % both on the anterior and posterior surface which is indicative of a homogenous water content. The parameter extraction is consistent across the scan range. On the right-hand side (b,d,f,h), phantom 7 is normalized with the 7.0 mm sphere. The data extraction is not consistent across the z-range. Even though the PSO does not lead to the expected water gradient, it is at least much more well behaved when the phantom and sphere have about the same outer radius of curvature.

Finally, in Fig. 15, the box plots of the eleven phantoms are reported for two different extraction techniques. In one case the thickness was a parameter to extract, and in the other it was the value derived from the OCT phantom images (Fig. 15 (a)). The thickness was constrained to a range from 450 um to 750 um, whereas the anterior water content was left vary from 0 to 100 % and the posterior water content from the anterior value to 100 %. The phantom water content was extracted by combining EMT and SMT as in [12], where EMT is based on a two-material Bruggeman model and the water gradient is assumed to be linear. The material model comprises of free water (double Debye model [28]) and a second constituent of permittivity 2.9. As reported in [29], a two material model has been theorized based on the assumption that collagen and bound water are linked to each other and their interaction with submillimeter waves, described by permittivity, is similar. However, the authors acknowledge that the model can benefit from further study of this particular collagen hydrogel and potential, subsequent model modification with additional constituents and/or revised relaxation terms.

The anterior water content is expected to be the most stable estimation parameter as it is the parameter that affects the amplitude the most. In the first three plots (b-d), the thickness is assumed to be an unknown and therefore there are three parameters to extract: anterior water content, posterior water content and thickness. The thickness extractions seem to perform poorly as it never reaches the true value, regardless of the phantom or the calibration measurement. The anterior water content and posterior water content values tend to overlap. Therefore, it is deduced that the phantoms have a homogenous water content as a function of depth. In our model drier phantoms should have a more oscillating frequency response if there is no gradient, but a gradient is highly likely to exist.

In the second group (e-f) the thickness was assumed to be known, and the thickness extracted with the OCT was used for the PSO fit. In this case, anterior and posterior water content might be different for most of the phantoms. This might be interpreted as the presence of a gradient or that the fit tends to lie on the boundaries and, therefore, there is no obvious solution. In [12] it was observed that the gradient can confound the thickness and posterior water content if the posterior water content is high, as the corresponding complex reflectivity becomes indistinguishable. As demonstrated by the quasioptical system and data analysis techniques tend to conserve optical path length; there is a strong negative correlation between extracted permittivity and thickness when thickness is treated as a free parameter. This same behaviour is seen in Fig. 15 (b), (c), and (d) where the increasing/decreasing thickness is paired with decreasing/increasing water content which is positively correlated with permittivity. Conversely, when the CCT is fixed, the optimization routine compensates by varying the permittivity. Since the anterior segment water content is primarily defined by the reflection coefficient amplitude, optical path length conservation is dependent mostly on the posterior content water content thus the large IQR in Fig. 15(f). The posterior water content fitting thus behaves poorly as the estimated PWC tends to land on the extrema; either 100% or the same as the AWC value thus suggesting no gradient. This behavior is likely indicative of unsuccessful fitting.

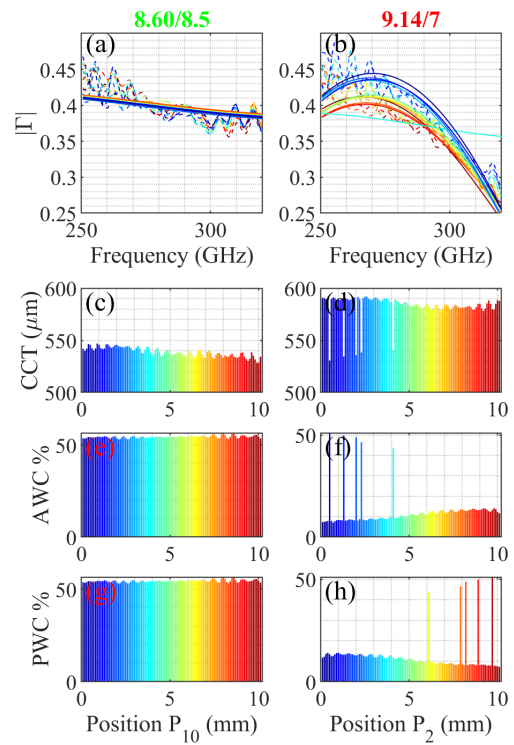

Fig. 14 (a) Phantom 25 measured (dashed lines) and fitted (solid lines) reflection coefficient in ten different positions when the calibration target RoC is 8.5 mm. (c) Extracted thickness (e) anterior water content and (g) posterior water content of phantom 25 along the z-scan when the calibration target RoC is 8.5 mm. (a) Phantom 7 measured (dashed lines) and fitted (solid lines) reflection coefficient in ten different positions when the calibration target RoC is 9.0 mm. (c) Extracted thickness (e) anterior water content and (g) posterior water content of phantom 7 along the z-scan when the calibration target RoC is 9.0 mm.

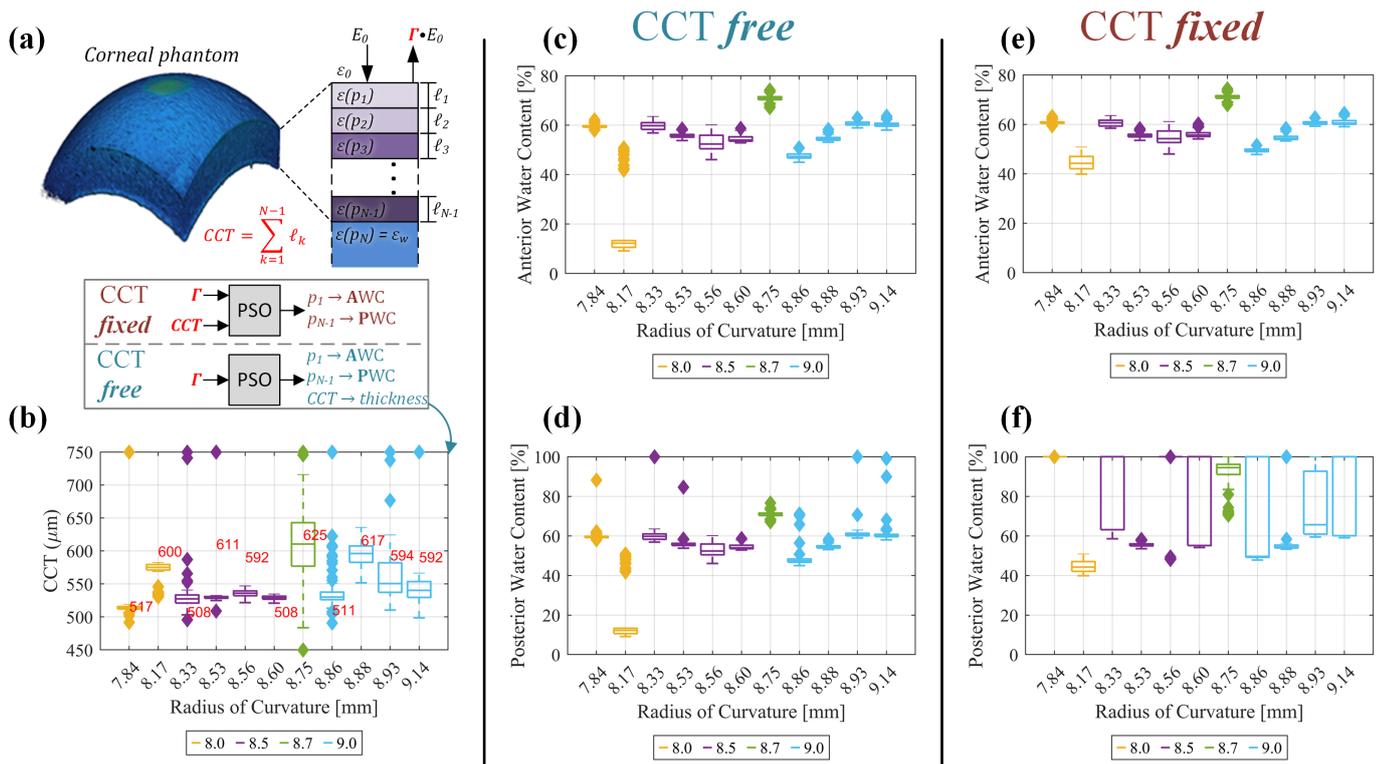

**Fig. 15** Box plots of the estimated phantom parameters where the x-axis reports the radius of the phantoms in mm. Each box plot represents data normalized with the closest steel sphere calibration target in term of RoC (section III). The calibration sphere radius of curvature (8.0, 8.5, 8.7, and 9.0 mm) is encoded in color and described by the legends in panels (b - f). (a) description of fitting routine. (b) Measured and phantom thickness. The measured values from OCT are reported in red text and overlap the box plot data which report the estimated thickness statistics extracted from millimeter wave data. (c) Extracted anterior water content with the CCT as free parameter (d) Extracted posterior water content with the CCT as free parameter. (e) Extracted anterior water content and (f) posterior water content with the OCT thickness measurements used as an input to the model.

## V. DISCUSSION

The axicon's sensitivity to target radius of curvature makes permittivity and thickness estimation from measured data challenging. The sensitivity arises from a redirection of the Bessel beam's first side lobes to the main lobe and the degree of redirection appears to be a strong function of RoC. The effects of these confounders can be mitigated when the target's geometry is well known and static, and there is a good match between target and calibration sphere RoC. The permittivity and thickness extraction from the air-backed quartz dome measurements across a broad axial range agree well with the known actual parameters.

In contrast, the gelatin phantom geometry was dynamic and difficult to control. The surface curvature was not quite spherical as evidenced by the OCT scans and over time the apex flattened out, increasing the central RoC. Thus, Bessel beam main lobe and side lobes were incident upon slightly different RoCs, and the RoC discrepancy increased over the total data acquisition time. These issues were further exacerbated by imperfect phantom and calibration sphere RoC matching and long experiment duration needed to accommodate the time-consuming peak search procedures.

Potential steps to address the methodological limitations include integrating a companion optical/imaging modality (e.g. OCT or THz) for online alignment verification and adopting a closed loop phantom pressure management system to minimize phantom geometry variability. Additionally, further development of a more mechanically robust phantom material is warranted, and a better understanding of the material dielectric properties will aid in fitting.

Further, smaller cone angles ($\alpha$ in [23]) will produce larger main lobes and thus sides lobes that miss the target completely at the expense of reduced coupling efficiency.

## VI. CONCLUSIONS

Submillimeter-wave reflectometry of a corneal phantom was carried out with AGBB. The aim was to extract the phantom water content and thickness by resolving the inverse scattering problem. PO Simulations, confirmed by experimental data, showed that the coupling between the AGBB and a sphere of radius with dimensions of ~7-11 mm is a strong function of the RoC. This made reflectometry calibration ambiguous for targets, which have a large radius of curvature. For targets, whose radii of curvature exceed 9 mm, the alignment with the optics becomes non-obvious, as the maximum of the reflection coefficient of a target moving along a transverse plane is no longer unique and does not happen when the target is located on the optical axis. The corneal phantom parameter extraction produces results that somewhat diverge from our previous publications and our understanding of the phantoms. Further refinements to the gelatin dielectric mode, reduced measurement time, and modifications of the axicon-hyperbolic lens design may improve the overall performance. Further, the submillimeter wave parameter estimation uncertainties for the quartz domes (thickness – 3.6 %, permittivity – 5.5 %) and gelatin phantoms (thickness – 9.2%, refractive index – 15.3%) for, suggest that additional work is needed to reduce measurement error prior to employing these techniques in a clinical setting.

ACKNOWLEDGMENTS

The authors would like to thank the Academy of Finland and the Icare Finland Oy for their financial support, Faezeh Zarrinkhat for her help with the measurements, Pyry Kiviharju for his implementation of the OCT segmentation code, Arthur Aspelin for the development of the PI stage moving code, and Dr. Mirmoosa for the highly valuable discussions.

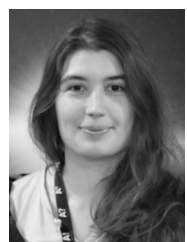

**Mariangela Baggio** (Student Member IEEE) was born in Camposampiero, Italy in 1992. She received her B. Sc. degree in Information Technology Engineering and M. Sc. degree in Telecommunication Engineering from the University of Padova, Italy in 2015 and 2018, respectively. She is currently pursuing her Ph.D. in Electrical Engineering at Aalto University, Finland. Her research interests include mmw and THz techniques, quasioptical system design and THz sensing and imaging.

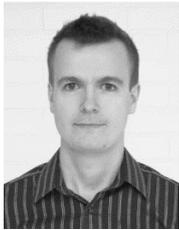

**Aleksi Tamminen** was born in Ruotsinpyhtää, Finland, in 1982. He received the B.Sc. (Tech.) and M.Sc. (Tech.) degrees from the Helsinki University of Technology, Espoo, Finland, in 2005 and 2007, respectively. He received the Lic.Sc. (Tech.) and D.Sc. (Tech.) degrees from Aalto University (former Helsinki University of Technology), Espoo, in 2011 and 2013, respectively. From 2005 to 2013, he was with the Department of Radio Science and Engineering, Aalto University. His research work in Aalto University was related to antenna measurements and imaging at millimeter and submillimeter waves. In addition to research, he served as the organizing secretary of international conference, 6th ESA Workshop on Millimetre-Wave Technology and Applications and 4th Global Symposium on Millimeter Waves, held in Espoo in May 2011. He has authored or co-authored about 50 scientific journal and conference publications as well as three patent applications. He received the Young Engineer Prize from the 5th European Radar Conference on 31st November 2008 as well as the Best Student Paper Award from the Global Symposium on Millimeter Waves 2010 on 14th –16th April 2010. From 2013 to 2018 he was Research Scientist at Asqella Ltd., Helsinki, Finland. He was the principal in the research and development of commercial submillimeter-wave imaging technology as well as in participating in academic research projects in the related field. From 2018, he is with Aalto University as Research Fellow. His current research interests are submillimeter- and millimeter-wave projects including antenna measurements, sensing biological tissues, and imaging.

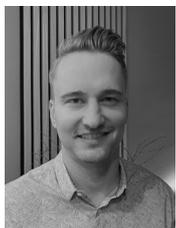

**Joel Lamberg** (Student Member, IEEE) was born in Helsinki, Finland, in 1985. He received a B.Sc. degree in electronics and electrical engineering and an M.Sc. degree in microwave engineering from Aalto University, Espoo, Finland, in 2019 and 2021, respectively. He is working toward a Ph.D. in radio science at Aalto University, Espoo, Finland. His research interests include electromagnetic modeling, terahertz imaging of human cornea, and quasi-optical system design.

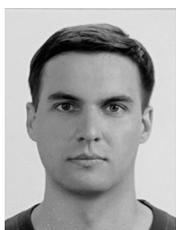

**Roman Grigorev** was born in Saint Petersburg, Russia in 1995. He received his M. Sc. degree in Photonics in 2018 from ITMO University, Russia, where he continued studies as a doctoral student in Optics. During his PhD course he worked at Seoul National University, South Korea as a research associate, and at Aalto University, Finland as a visiting doctoral candidate. His current research interests include terahertz technologies for biomedical diagnostic applications.

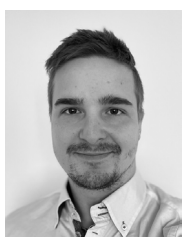

**Samu-Ville Pälli** (Student Member, IEEE) was born in Savonlinna, Finland, in 1994. He received the B.Sc. (Tech.) and M.Sc. (Tech.) degrees (Hons.) in electrical engineering from Aalto University, Espoo, Finland, in 2018 and 2020, respectively, where he is currently pursuing the D.Sc. (Tech.) degree. He has been with the Department of Electronics and Nanoengineering, School of Electrical Engineering, Aalto University, since 2017. His current research interests include computational imaging methods and systems at submillimeter waves and terahertz imaging and sensing.

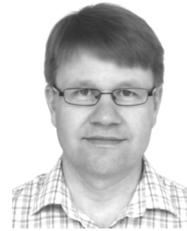

**Juha Ala-Laurinaho** received the Diploma Engineer (M.Sc.) degree in mathematics and D.Sc. (Tech.) degree in electrical engineering from TKK Helsinki University of Technology, Finland, in 1995 and 2001, respectively. He has been with the TKK, currently Aalto University, serving in the Radio Laboratory in 1995–2007, in the Department of Radio Science and Engineering in 2008-2016, and currently in the Department of Electronics and Nanoengineering. Currently, he works as a Staff Scientist. Dr. Ala-Laurinaho has been a Researcher and Project Manager in various millimeter wave technology related projects. His current research interests are the antennas and antenna measurement techniques for millimeter and submillimeter waves, and the millimeter wave imaging.

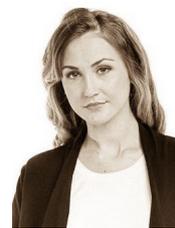

**Irina I. Nefedova** received her the B.Sc. and M.Sc. degrees in physics from Saratov State University, Saratov, Russia, in 2010 and 2012, respectively, and PhD. degree in electrical engineering from Aalto University, Finland, in 2017. She was a Postdoctoral Researcher and a Visiting Scientist with the University of Hamburg, in 2018, the Jet Propulsion Laboratory, California Institute of Technology (Caltech), in 2019, and the Chalmers University of Technology, in 2020–2021. Since 2019, she has been a Postdoctoral Researcher at Aalto University. Her interests include millimeter-wave, terahertz, and quasi-optical measurements with applications to remote sensing and medical imaging.

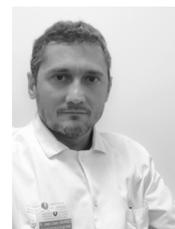

**Pr Jean-Louis Bourges** achieved a medical residency and a fellowship of ophthalmology in Paris university. He completed an international fellowship from the Jules Stein Eye Institute at David Geffen school of medicine, UCLA. He is MD, PhD, and fellow of the European board of ophthalmology. He serves as full professor of French universities at the *Université Paris Cité*, Paris Descartes school of medicine. He is in charge of surgical pedagogy and training by simulation at the French college of professors in ophthalmology. His clinical practice at Cochin Hospital is dedicated to ophthalmology, and subspecialized in anterior segment surgery, cataract, refractive surgery and corneal diseases. He supervises eye emergency at *Assistance Publique-Hôpitaux de Paris* and the corneal division of the ophtalmopole, APHP, Paris. He published more than 75 scientific peer-reviewed papers, 20 book chapters and other online publications in the field of ophthalmology. He edited three books about cornea and ocular emergency. He is also medical expert in the field of aeronautic ophthalmology.

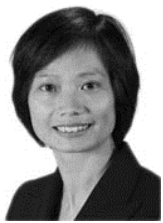

**Sophie X. Deng** received the B.S. degree from the City College of New York in 1991. In 2001, Deng received her doctor of medicine and doctor of philosophy degrees at the University of Rochester School of Medicine and Dentistry, New York, where she completed the rigorous Medical Scientist Training Program. She is an Associate Professor in the Cornea Division at the Jules Stein Eye Institute, UCLA. She did her residency in ophthalmology at the Illinois Eye and Ear Infirmary, Chicago. Dr. Deng subsequently completed her fellowship in Cornea, External Ocular Disease and Refractive Surgery at the Jules Stein Eye Institute. Dr. Deng is a specialist in corneal and external ocular diseases, and cataracts. Her surgical areas include endothelial keratoplasty (DSEK and DMEK), deep anterior lamellar keratoplasty (DALK), penetrating keratoplasty, limbal stem cell transplantation, artificial cornea and cataract. Dr. Deng is the director of the Cornea Biology Laboratory at the Jules Stein Eye Institute. Her research focuses on corneal epithelial stem cells regulation, deficiency and regeneration. Dr. Deng's research aims to improve the current treatment for patients with limbal stem cell deficiency by using stem cell therapy to restore vision. In addition, Dr. Deng conducts clinical studies to develop new imaging and molecular tests to accurately diagnose and stage limbal stem cell deficiency.

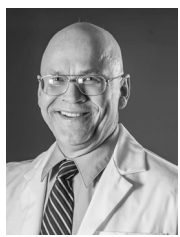

**Elliott R. Brown** (Fellow IEEE) received the Ph.D. and M.S. degrees in applied physics from the California Institute of Technology, Pasadena, CA, USA in 1985 and 1981, respectively. Currently, he is a Professor (Emeritus) of physics and electrical engineering at Wright State University (WSU), Dayton OH, USA. Previously he was a Professor of EE at the University of California, (Santa Barbara and Los Angeles campuses), and prior to that was a Program Manager at DARPA in Arlington, VA, and a research-group leader at MIT Lincoln Laboratory. He has taught courses in sensor physics, electromagnetics, THz science and technology, solid-state physics, and microfabrication science. His research encompasses mm-wave-to-THz ultrasensitive room temperature detectors; GaN and InGaAs resonant tunneling and light emitting devices; ultrafast 1550-nm GaAs photoconductive and quantum dot sources; and mm-wave-to-THz biomedical imaging and inverse synthetic aperture radar. He is a Fellow of the IEEE (since 2000), the American Physical Society (since 2007), and the Optical Society of America (since 2020). In 2010 he was selected for the Ohio Research Scholars Endowed Chair in THz Sensors Physics. In 2016 he was selected for the Chair-of-Excellence at the University Carlos III in Madrid Spain, and while in Europe established several collaborations that remain fruitful to the present.

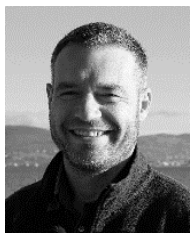

**Vincent Wallace** is an Associate Professor of Physics at the University of Western Australia, Perth, Australia. He graduated with a PhD in Physics from the University of London, in 1997. After three years at the Beckman Laser Institute, California, developing multiphoton imaging he joined Toshiba Research and subsequently TeraView Ltd in Cambridge, UK to develop biomedical applications of Terahertz Technology. In 2007 he moved to the University of Western Australia where he continues to work in terahertz and optical technologies.

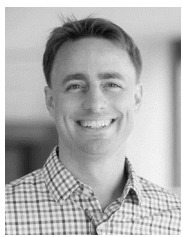

**Zachary D. Taylor** (S'06–M'09) received the B.S. degree in electrical engineering from the University of California (UCLA), Los Angeles, CA, USA, in 2004, and the M.S. and Ph.D. degrees in electrical engineering from the University of California (UCSB), Santa Barbara, CA, USA, in 2006 and 2009, respectively. From 2013 – 2018 he was an adjunct assistant professor with appointments in the Department of Bioengineering, Department of Electrical Engineering, and Department of Surgery at the University of California (UCLA), Los Angeles, CA, USA. From 2018-2022, he was an assistant professor in the Department of Electronics and Nanoengineering at Aalto University, Espoo, Finland. From 2022, he is associate professor in the same department. His current research interests include submillimeter-wave and THz imaging and sensing and THz frequency calibration techniques for antenna measurements, personnel imaging, and clinical diagnostics.